\documentclass{article}

\usepackage{microtype}
\usepackage{graphicx}
\usepackage{subcaption}
\usepackage{booktabs} % for professional tables

\usepackage{hyperref}

\usepackage[accepted]{icml2026}

\usepackage{amsmath}
\usepackage{amssymb}
\usepackage{mathtools}
\usepackage{amsthm}
\usepackage{braket}

\usepackage[capitalize,noabbrev]{cleveref}

\theoremstyle{plain}

\theoremstyle{definition}

\theoremstyle{remark}

\newcommand{\normdistvae}{\mathrm{norm\_dist}_{\mathrm{VAE}}}
\newcommand{\normdistq}{\mathrm{norm\_dist}_{\mathrm{Q}}}
\newcommand{\normdir}{\mathrm{norm\_dir}}
\icmltitlerunning{Q-BAR: Blogger Anomaly Recognition}

\begin{document}

\twocolumn[
  \icmltitle{Q-BAR: Blogger Anomaly Recognition via \\Quantum-enhanced Manifold Learning}

  \icmlsetsymbol{equal}{*}

  \begin{icmlauthorlist}
    \icmlauthor{Maida Wang}{yyy}
    \icmlauthor{Panyun Jiang}{xxx}
  \end{icmlauthorlist}

  \icmlaffiliation{yyy}{Centre for Computational Science, University College London, London, UK}
  \icmlaffiliation{xxx}{Institute of Education, University College London, London, UK}

  \icmlcorrespondingauthor{Maida Wang}{maida.wang.24@ucl.ac.uk}

  \icmlkeywords{Video Anomaly Detection, Multimodal Learning, Misinformation}

  \vskip 0.3in
]

\printAffiliationsAndNotice{}  % no special notice (required even if empty)

\begin{abstract}
In recommendation-driven online media, creators increasingly suffer from semantic mutation, where malicious secondary edits preserve visual fidelity while altering the intended meaning. Detecting these mutations requires modeling a creator's unique semantic manifold. However, training robust detector models for individual creators is challenged by data scarcity, as a distinct blogger may typically have fewer than 50 representative samples available for training. We propose quantum-enhanced blogger anomaly recognition (Q-BAR), a hybrid quantum-classical framework that leverages the high expressivity and parameter efficiency of variational quantum circuits to detect semantic anomalies in low-data regimes. Unlike classical deep anomaly detectors that often struggle to generalize from sparse data, our method employs a parameter-efficient quantum anomaly detection strategy to map multimodal features into a Hilbert space hypersphere. On a curated dataset of 100 creators, our quantum-enhanced approach achieves robust detection performance with significantly fewer trainable parameters compared to classical baselines. By utilizing only hundreds of quantum parameters, the model effectively mitigates overfitting, demonstrating the potential of quantum machine learning for personalized media forensics.
\end{abstract}

\section{Introduction}
\label{sec:intro}

The digital information ecosystem is currently witnessing a crisis of authenticity that transcends pixel-level forgery~\cite{farid2022creating, wardle2017information}. While deepfake detection focuses on AI-synthesized artifacts, a more pervasive threat has emerged in the form of semantic mutation~\cite{guarnera2020fighting}. In this phenomenon, unscrupulous actors, often operating as content farms, use traditional editing tools to reorder, decontextualize, or selectively splice genuine footage~\cite{mubarak2023survey}. A video subjected to these attacks remains visually authentic, containing the real speaker and genuine audio, but the underlying semantics and intent are fundamentally mutated to fabricate false narratives or clickbait~\cite{kishwar2025regulating,luo2021newsclippings}.

The societal impact of such manipulation is profound and immediate
~\cite{vaccari2020deepfakes}. A prominent example occurred on short-video platforms involving the top-tier technology creator \textit{MediaStorm} (Yingshijufeng)\footnote{For example, the widely discussed incident regarding unauthorized re-editing and platform content degradation reported in \textit{https://www.zhihu.com/question/1972852612236850378} (2025) or MediaStorm's official response on Bilibili.}. Malicious re-editing of their reviews resulted in a complete semantic inversion of the original critique, fabricating conflicts that did not exist. This incident not only dominated platform trending lists but also instigated widespread misinformation and targeted cyber-harassment against the creator. Such events underscore a critical vulnerability in the current media landscape: when visual fidelity is weaponized to distort factual truth, it can rapidly incite irrational public sentiment and cause irreversible reputational damage before the original context can be restored~\cite{vosoughi2018spread, vaccari2020deepfakes}.
Detecting these high-level logic violations requires modeling the unique semantic manifold of the original creator~\cite{ruff2018deep}. A genuine blogger maintains a consistent latent signature defined by their linguistic patterns, logical structures, and prosodic delivery~\cite{stamatatos2009survey}. Ideally, an anomaly detector would learn the topology of this manifold and flag any video that drifts into low-density regions, indicating a deviation from the creator's established persona.

\begin{figure*}[htbp]
    \centering
    \includegraphics[width=0.9\textwidth]{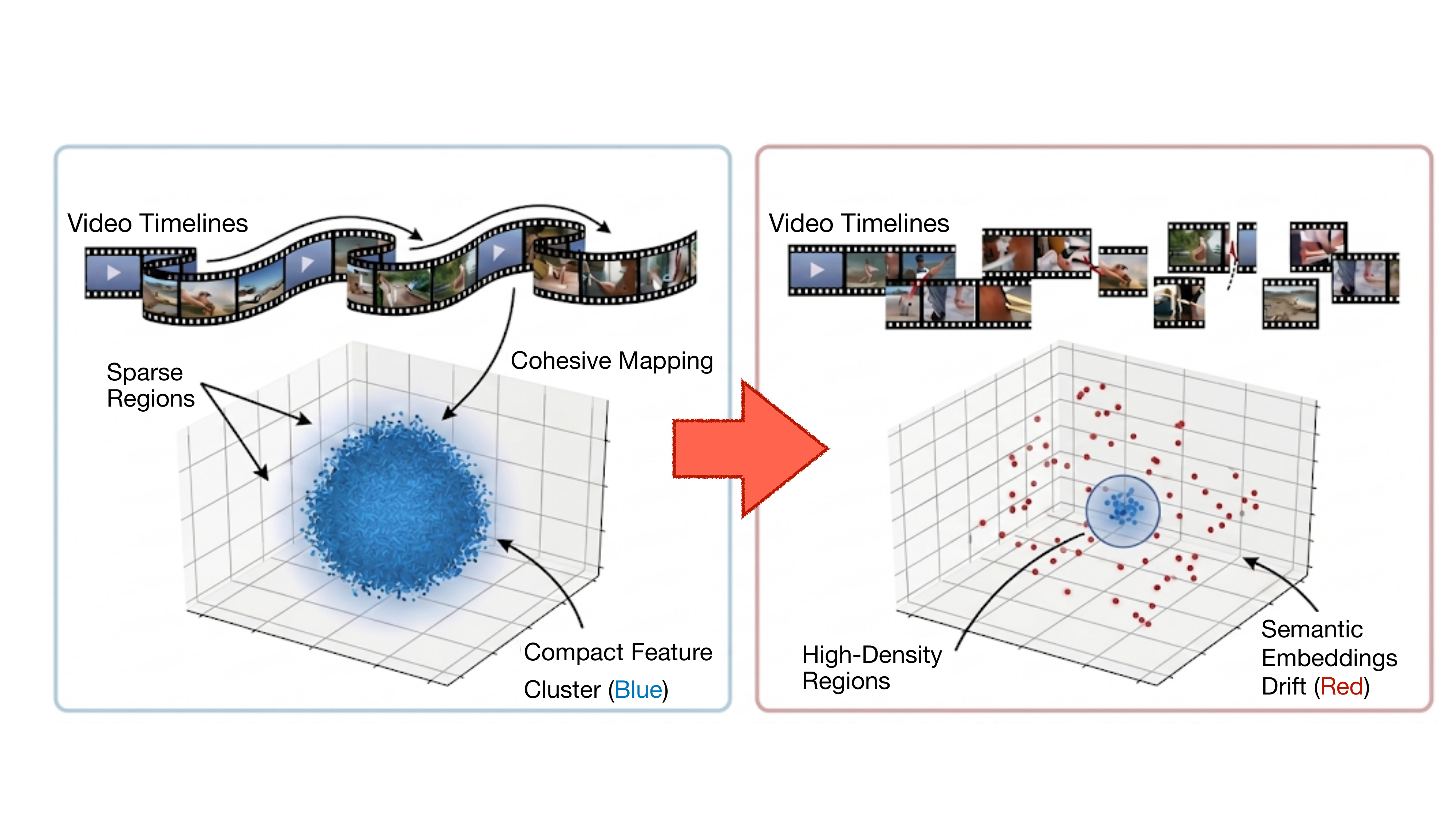}
    \caption[\textbf{Conceptual Illustration of Semantic Mutation}.]{Conceptual Illustration of Semantic Mutation. The figure contrasts the latent topological structure of authentic versus manipulated content. (Left) A genuine creator maintains a cohesive ``Semantic Manifold,'' represented by continuous video timelines and a compact, high-density feature cluster (blue). (Right) Malicious marketing accounts employ fragmentation and splicing, causing the semantic embeddings to drift into low-density regions (red), a phenomenon we define as ``Semantic Mutation.''}
    \label{fig:concept_comparison}
\end{figure*}

However, applying classical deep learning to this task faces a critical bottleneck: data scarcity~\cite{huang2022quantum,huang2021power,wang2025parameter}. A specific creator typically possesses a limited repository of high-quality, long-form original videos, often ranging from 20 to 50 samples.  In our experiments, we further restrict the training set to approximately 20 videos per creator in order to simulate an extreme low-data regime,  which reflects realistic deployment scenarios where only a small number of verified original videos are available. Classical anomaly detection methods, such as Deep Support Vector Data Description (Deep SVDD)~\cite{ruff2018deep} or Variational Autoencoders (VAE)~\cite{park2023variational, an2015variational}, are notoriously data-hungry. When trained on such small, high-dimensional multimodal datasets, these models tend to suffer from severe overfitting or require massive parameter sets that fail to generalize to unseen semantic attacks~\cite{wang2023quantum}.
To surmount the obstacle of data scarcity, we turn to the emerging paradigm of Quantum Machine Learning (QML)~\cite{biamonte2017quantum}. Recent theoretical breakthroughs have rigorously established that quantum models can extract physical information from complex, high-dimensional feature spaces more efficiently than their classical counterparts~\cite{wang2025quantum}. Notably, Huang et al. demonstrated that quantum learning algorithms can achieve a provable advantage in sample complexity, enabling accurate classification with significantly fewer training examples~\cite{huang2021power, huang2022quantum}. This characteristic renders QML uniquely suitable for modeling the semantic manifolds of individual creators, where the ``ground truth'' is sparse yet highly complex.

Motivated by these findings, our approach integrates the Parameter-Efficient Quantum Anomaly Detection (PEQAD) algorithm into the domain of multimedia forensics~\cite{wang2025parameter}. Specifically, we fuse textual, audio, and visual features into a compact representation and employ a variational quantum circuit (VQC) to map normal blogger content into a minimal hypersphere in the quantum state space. This strategic integration yields a threefold enhancement. First, it mitigates overfitting by achieving robust anomaly detection using a few trainable quantum parameters, whereas comparable classical models often require hundreds. Second, the quantum state representation naturally models the entangled relationships between a blogger's tone and logic~\cite{gao2022enhancing}. Third, the reduced model complexity facilitates potential deployment on near-term quantum devices or efficient simulators.

To the best of our knowledge, this work represents the first application of anomaly detection to the problem of semantic integrity and misinformation. We formalize the problem of semantic mutation, propose a manifold learning framework based on quantum state mapping, and demonstrate that quantum-enhanced models can perform well in creator-specific, low-resource scenarios.

\section{Related Work}

\paragraph{Video Anomaly Detection (VAD).}
Traditional VAD identifies abnormal events in surveillance feeds, typically employing autoencoders~\cite{hasan2016learning} or multiple instance learning to detect deviations from normal motion patterns~\cite{sultani2018real}. However, these methods rely heavily on visual cues (optical flow, gradients)~\cite{liu2018future} and are agnostic to linguistic semantics. They do not focus on detecting a video where a person is speaking calmly but saying something successfully spliced to reverse their original meaning~\cite{abdalla2025video}.

\paragraph{Misinformation and Deepfake Detection.}
Research in misinformation has largely focused on ``Fake News'' classification via textual features or graph-based propagation patterns. Meanwhile, Deepfake detection analyzes frequency domain artefacts or blending boundaries in synthesized faces \cite{rossler2019faceforensics}. Our work addresses a distinct category: \textit{Cheapfakes} or \textit{shallowfakes}, where the manipulation is editorial rather than generative. As noted by \cite{paris2019deepfakes}, this semantic manipulation is often more dangerous and harder to detect than AI-generated content. Crucially, existing deepfake detectors fail on these semantic alterations because no pixel-level artefacts are present.

\paragraph{Semantic Consistency.}
Recent works in NLP have explored logical consistency in text summarization \cite{maynez2020faithfulness}. We extend this notion to the multimodal domain, treating the blogger's historical corpus as the ``ground truth'' reference for consistency checking.

\paragraph{Quantum Machine Learning.}
Quantum Machine Learning explores the integration of quantum computing principles with classical learning algorithms to address complex computational challenges. Recent theoretical breakthroughs have demonstrated that QML models can possess a significant advantage in sample complexity compared to classical counterparts \cite{huang2021power, huang2022quantum}. Specifically, quantum models can extract effective information from high-dimensional feature spaces using significantly fewer training samples. This property is particularly relevant to our task of modeling individual creators, where the available "ground truth" data is inherently scarce and high-dimensional, making classical deep learning prone to overfitting.

\paragraph{Quantum Anomaly Detection and PEQAD.}
While anomaly detection is a well-established field for identifying deviations from a learned norm~\cite{lavin2015evaluating3, garcia2009anomaly4, phua2010comprehensive5, schlegl2017unsupervised6}, the application of QML to this domain remains nascent. Existing quantum approaches, such as kernel-based methods~\cite{belis2024quantum} or the Variational Quantum One-Class Classifier (VQOCC)~\cite{park2023variational}, often struggle with hardware efficiency or lack experimental validation on general image datasets. To address these limitations, we adopt the PEQAD framework. Unlike prior methods, PEQAD employs an end-to-end variational circuit that maps multimodal features into a minimal hypersphere in the Hilbert space. Crucially, theoretical analysis and empirical results on superconducting processors suggest that PEQAD can achieve superior expressivity and detection performance with orders of magnitude fewer parameters than classical Deep SVDD~\cite{ruff2018deep}. This extreme parameter efficiency makes it uniquely suited for detecting subtle semantic mutations in low-resource environments.

\section{Methodology}
\label{sec:method}

Semantic representations of multimedia content often exhibit complex cross-modal correlations. 
For instance, the meaning of a statement may depend jointly on textual phrasing, vocal prosody, and visual context. 
Capturing such interactions using classical models typically requires high-dimensional feature expansions or deep neural networks with large numbers of parameters.
Quantum feature maps provide an alternative mechanism for representing such correlations. 
By embedding classical feature vectors into a quantum Hilbert space through parametrized unitary transformations, variational quantum circuits implicitly generate highly nonlinear feature maps whose effective dimension grows exponentially with the number of qubits. 
Although the circuit itself may contain only a small number of trainable parameters, the resulting state representation can capture complex interactions among input features.
In the context of semantic mutation detection, this property is particularly desirable. 
Subtle manipulations such as sentence reordering, contextual deletion, or mismatches between tone and content may correspond to high-order correlations between modalities rather than simple feature deviations. 
Quantum embeddings provide a compact mechanism to model these interactions while maintaining a small parameter footprint, making them suitable for creator-specific learning under limited data.

Our framework is predicated on the hypothesis that a genuine blogger maintains a consistent ``Semantic Manifold''---a latent high-dimensional space defined by their historical linguistic patterns, logical structures, and stance stability. For instance, consider a creator who consistently expresses support for a specific sports team (e.g., the French football team) across years of content. These videos will cluster tightly within a specific high-density region of the semantic space. If a malicious actor splices footage to fabricate a narrative of hostility toward that same team, the resulting feature vector will drift significantly away from this established cluster, effectively ''falling off" the manifold.

Formally, let $\mathcal{M}_c \subset \mathbb{R}^D$ denote the semantic manifold associated with a specific creator $c$. 
Each authentic video $v_i$ is mapped to a multimodal feature vector $x_i \in \mathbb{R}^D$, and the set of historical content 
$\{x_1, \dots, x_N\}$ forms a compact region approximating $\mathcal{M}_c$. 
Malicious edits introduce semantic perturbations that cause the embedding $\hat{x}$ of a manipulated video $\hat{v}$ to deviate 
from this region, producing a measurable anomaly signal.

\begin{figure*}[ht]
    \centering
    % 确保文件名和路径正确
    \includegraphics[width=0.85\textwidth]{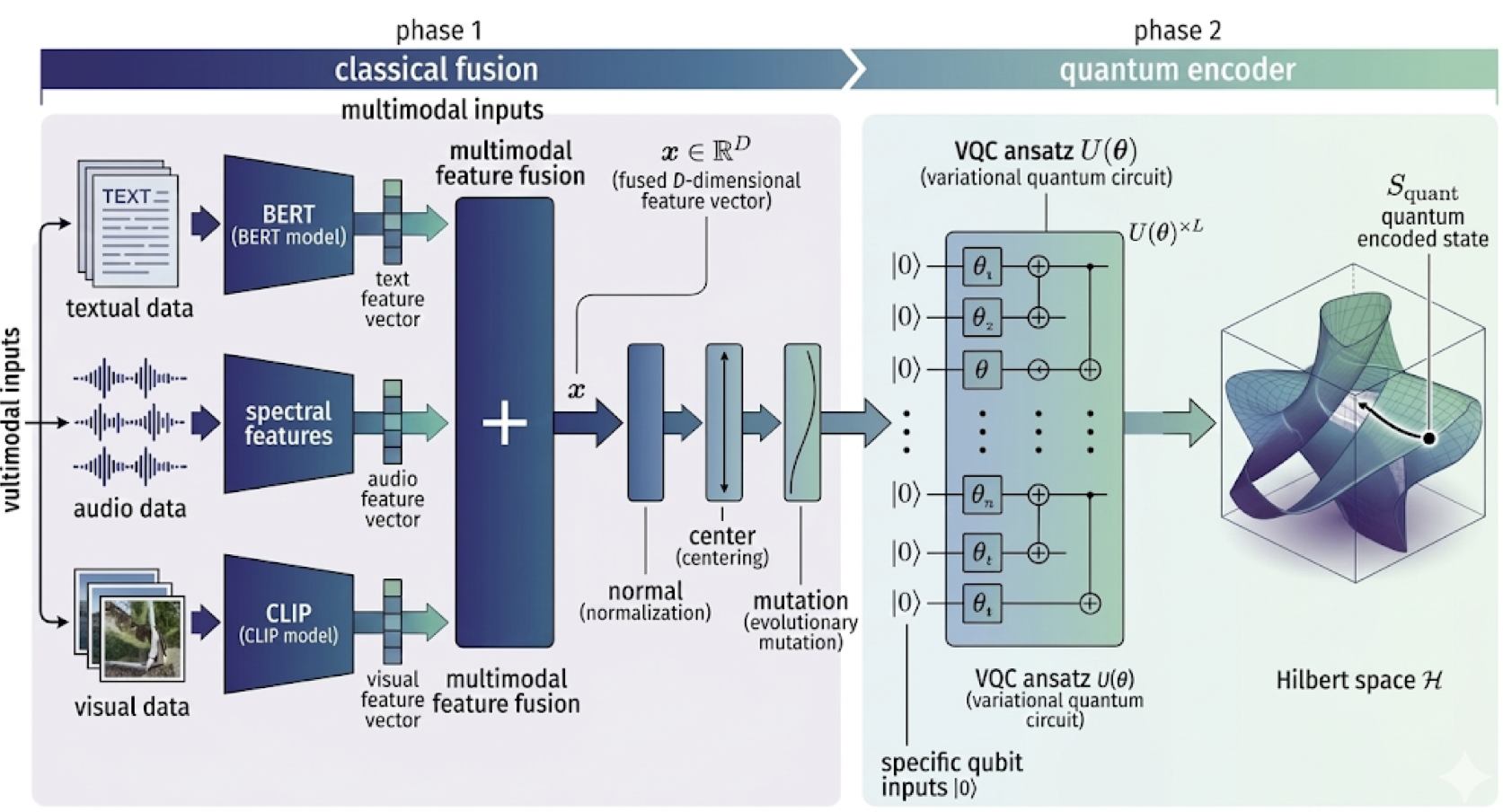}
    \caption{{Overview of the Quantum-BAR Framework.} 
    (1) \textbf{Classical Fusion:} Textual, audio, and visual features are extracted and fused into a vector $x$. 
    (2) \textbf{Quantum Encoder:} The vector is amplitude-encoded into a 12-qubit system and processed by a VQC with parameters $\theta$. 
    (3) \textbf{Semantic Manifold:} The VQC maps normal videos (blue dots) into a compact hypersphere centered at $|c\rangle$. Malicious mutations drift away from the center ($S_{quant}$) and align with the sensationalism axis ($S_{dir}$).}
    \label{fig:architecture}
\end{figure*}

To detect such mutations, we require a mapping function that is sensitive to these subtle high-dimensional shifts but robust to normal variations. We achieve this by projecting multimodal features into a quantum Hilbert space using the PEQAD framework, as shown in Fig.~\ref{fig:architecture}.

\subsection{Multimodal Feature Extraction and Fusion}
We represent each video \(v\) by a set of modality-specific embeddings \( \{f_{text}(v), f_{audio}(v), f_{visual}(v), f_{meta}(v)\} \). To construct a comprehensive semantic representation, we extract features across four distinct modalities. Textual features are derived from Automatic Speech Recognition (ASR) transcripts, which are processed using a pre-trained BERT encoder~\cite{gardazi2025bert} to capture the underlying semantic logic and linguistic patterns. Simultaneously, visual context is encoded through frame-level embeddings obtained via the CLIP model \cite{radford2021learning}, ensuring that visual fidelity is aligned with semantic content. Audio features incorporate prosodic elements, such as pitch and jitter, alongside spectral statistics; these are critical for identifying synthetic artifacts introduced by Text-to-Speech (TTS) systems, often used in malicious edits. Finally, we incorporate metadata features, including posting timestamps and tag usage distributions, to provide temporal and categorical context.

Before fusion, we apply layer normalisation to each modality vector $\tilde f_m(v)$. These are then concatenated into a single high-dimensional semantic vector $x \in \mathbb{R}^{D}$:
\begin{equation}
x = \mathrm{concat}\big(f_{text}^{(norm)}, f_{audio}^{(norm)}, f_{visual}^{(norm)}, f_{meta}^{(norm)}\big).
\end{equation}
Given a 12-qubit quantum system, our feature space dimension is $D_q = 2^{12} = 4096$. If the concatenated dimension $D < D_q$, we pad $x$ with zeros; if $D > D_q$, we apply a linear projection layer $W_p$ to compress the features to $D_q$.

\subsection{Quantum Manifold Modeling via PEQAD}
To model the creator-specific semantic manifold, we move beyond classical autoencoders, which often overfit on small creator-specific datasets (typically $<50$ videos). Instead, we employ a VQC to map the classical vector $x$ into a quantum state $|\psi(x)\rangle$.

\subsubsection{Amplitude Encoding}
We utilize \emph{Amplitude Encoding} to map the normalized classical vector $x$ into the amplitudes of a 12-qubit quantum state. This method allows us to encode $2^n$ classical features into $n$ qubits, providing compression and capturing high-dimensional correlations that are computationally expensive for classical kernels \cite{wang2025parameter}, although efficient state preparation remains an open challenge. The encoded state is:
\begin{equation}
|\psi_{in}(x)\rangle = \sum_{i=0}^{2^{12}-1} x_i |i\rangle, \quad \text{where } \sum |x_i|^2 = 1.
\end{equation}
Our use of Amplitude Encoding is motivated by prior findings that it significantly outperforms Angle Encoding (Rotation Encoding) in capturing complex image and semantic structures[cite: 233, 234].

\subsubsection{Variational Quantum Circuit (VQC)}
The encoded state evolves through a parametrized unitary circuit $U(\theta)$, consisting of entangling layers (CNOT gates) and trainable single-qubit rotations ($R_y, R_z$). The circuit parameters $\theta$ define the topology of the learned manifold. The output state is given by:
\begin{equation}
|\psi_{out}(x;\theta)\rangle = U(\theta) |\psi_{in}(x)\rangle.
\end{equation}
We adopt a hardware-efficient ansatz similar to the "Real Amplitudes" or "Dressed Quantum Circuit" structures, which have been shown to maintain high expressivity with limited depth[cite: 441].

\subsubsection{Hypersphere Optimization Objective}
Following the Deep SVDD objective adapted for quantum circuits, we train the VQC to map all normal videos (the blogger's authentic history) into a minimum-volume hypersphere in the Hilbert space, centered at a fixed state $|c\rangle$. Following the PEQAD formulation \cite{wang2025parameter}, the center state $|c\rangle$ is defined as the reference state 
representing the normal semantic region of the creator. In practice, $|c\rangle$ is determined according to the 
initialization strategy described in PEQAD, which ensures that the hypersphere center remains fixed during training 
while the circuit parameters $\theta$ adapt to minimize the distance between encoded samples and the center state. The optimization objective is to minimize the expected distance between the mapped states and the center:
\begin{equation}
\min_\theta L_{PEQAD} =
\frac{1}{N}\sum_{i=1}^{N}
\left(1 - |\langle c | \psi_{out}(x_i; \theta) \rangle|^2 \right)
+ \frac{\lambda}{2}\|\theta\|_F^2
\end{equation}
where $N$ is the batch size and $\lambda$ is a regularization term. This formulation corresponds to maximizing the quantum state fidelity between encoded samples and the center state. By pulling all authentic semantic vectors toward $|c\rangle$, the VQC learns a compact manifold $\mathcal{M}$ where the creator's logic and style are consistent. 

\subsection{Anomaly Scoring and Inference}
For a new query video $\hat{v}$ (potentially a malicious edit), we compute its anomaly score based on two factors: deviation from the semantic manifold and alignment with sensationalist patterns.

\paragraph{Quantum Semantic Deviation ($S_{quant}$).}
We encode $\hat{v}$ into the quantum circuit and measure the distance of its output state from the learned center $|c\rangle$:
\begin{equation}
S_{quant}(\hat{v}) = \big\| |\psi_{out}(\hat{x};\theta^*)\rangle - |c\rangle \big\|^2.
\end{equation}
A high $S_{quant}$ indicates that the video's logic structure falls outside the creator's established semantic manifold.

\paragraph{Directional Mutation Score ($S_{dir}$).}
Semantic mutations often drift not just away from the creator, but towards clickbait or conflict. We learn a global ``Sensationalism Direction'' $w_{mut}$ using Linear Discriminant Analysis on a held-out corpus 
of clickbait headlines and rumor-related content collected from publicly available misinformation datasets. 
This direction captures semantic patterns commonly associated with sensationalized narratives. The directional score is the cosine similarity:
\begin{equation}
S_{dir}(\hat v) = \frac{\langle x, \mathbf{w}_{mut}\rangle}{\|\mathbf{w}_{mut}\| \,\|x\|}.
\end{equation}

\paragraph{Final Decision.}
The final anomaly score combines two complementary signals: structural deviation from the creator-specific semantic manifold and directional drift toward sensationalized content patterns. These two components capture different aspects of semantic mutation and are combined through a weighted linear score:
\begin{equation}
S_{final}(\hat v) = \gamma \cdot \mathcal{N}(S_{quant}) + (1-\gamma) \cdot \mathcal{N}(S_{dir}),
\end{equation}
where $\mathcal{N}(\cdot)$ denotes Z-score normalization and $\gamma$ is a weighting hyperparameter. If $S_{final} > \tau$, the video is flagged as a semantic anomaly.

\subsection{Training and Inference Procedure}

The complete Q-BAR pipeline is summarized in Algorithm~\ref{alg:qbar}. 
The procedure preserves the original BAR components, including multimodal fusion, classical manifold modelling, retrieval-based alignment, and directional mutation scoring, while augmenting them with the PEQAD quantum anomaly detector. 
In practice, the classical and quantum modules are trained separately and fused only at the score level, which improves stability in the low-data regime.

\begin{algorithm}[t]
\caption{Q-BAR: Quantum-enhanced Blogger Anomaly Recognition}
\label{alg:qbar}
\begin{algorithmic}[1]

\STATE \textbf{Input:} creator-specific authentic dataset $\mathcal{D}=\{v_i\}_{i=1}^{N}$, query clip $\hat v$, external clickbait corpus $\mathcal{C}$

\STATE \textbf{Multimodal feature extraction:}
\FOR{each video $v_i \in \mathcal{D}$}
    \STATE Extract textual, audio, visual, and metadata features
    \STATE Compute fused representation
    \STATE $f(v_i) \leftarrow \mathrm{Fuse}\!\left(f_{\text{text}}(v_i), f_{\text{audio}}(v_i), f_{\text{visual}}(v_i), f_{\text{meta}}(v_i)\right)$
\ENDFOR

\STATE \textbf{Classical manifold modelling:}
\STATE Train a VAE on $\{f(v_i)\}_{i=1}^{N}$ to obtain encoder $E$ and decoder $G$
\STATE Train a Deep SVDD model on $\{f(v_i)\}_{i=1}^{N}$ to estimate a compact support region

\STATE \textbf{Quantum manifold modelling:}
\STATE Reduce fused features to a quantum-compatible dimension: $x_i \leftarrow r(f(v_i))$
\STATE Initialise PEQAD variational quantum circuit $U_\theta$ and centre state $\ket{c}$

\FOR{each training sample $x_i$}
    \STATE Encode $x_i$ into an input quantum state $\ket{\psi_{\mathrm{in}}(x_i)}$
    \STATE Apply variational circuit to obtain $\ket{\psi_{\mathrm{out}}(x_i;\theta)} = U_\theta \ket{\psi_{\mathrm{in}}(x_i)}$
\ENDFOR

\STATE Optimise $\theta$ by minimizing
\[
\mathcal{L}_{Q}
=
\frac{1}{N}\sum_{i=1}^{N}
\left(1-\left|\langle c \mid \psi_{\mathrm{out}}(x_i;\theta)\rangle\right|^2\right)
+\frac{\lambda}{2}\|\theta\|_F^2
\]

\STATE \textbf{Directional mutation modelling:}
\STATE Learn the sensationalism direction $w_{\mathrm{mut}}$ from $\mathcal{C}$ using LDA

\STATE \textbf{Inference for query clip $\hat v$:}
\STATE Compute fused feature vector
\STATE $f(\hat v) \leftarrow \mathrm{Fuse}\!\left(f_{\text{text}}(\hat v), f_{\text{audio}}(\hat v), f_{\text{visual}}(\hat v), f_{\text{meta}}(\hat v)\right)$

\STATE Optionally retrieve candidate source videos using pHash and CLIP, then align them using DTW

\STATE Compute classical anomaly score
\STATE $S_{\mathrm{VAE}}(\hat v) \leftarrow \|f(\hat v)-G(E(f(\hat v)))\|_2^2$

\STATE Compute quantum anomaly score
\STATE $\hat x \leftarrow r(f(\hat v))$
\STATE Prepare $\ket{\psi_{\mathrm{out}}(\hat x;\theta)} = U_\theta \mathrm{Encode}(\hat x)$
\STATE $S_Q(\hat v) \leftarrow 1-\left|\langle c \mid \psi_{\mathrm{out}}(\hat x;\theta)\rangle\right|^2$

\STATE Compute directional mutation score
\STATE $S_{\mathrm{dir}}(\hat v) \leftarrow \langle f(\hat v), w_{\mathrm{mut}} \rangle / \left(\|f(\hat v)\|\,\|w_{\mathrm{mut}}\|\right)$

\STATE Normalize the three scores on the validation distribution:
\STATE $\normdistvae(\hat v)$, $\normdistq(\hat v)$, $\normdir(\hat v)$

\STATE Compute final anomaly score
\[
S(\hat v)
=
\beta_1 \normdistvae(\hat v)
+
\beta_2 \normdistq(\hat v)
+
(1-\beta_1-\beta_2)\normdir(\hat v)
\]

\IF{$S(\hat v) > \tau$}
    \STATE Flag $\hat v$ as a semantic anomaly
\ELSE
    \STATE Treat $\hat v$ as consistent with the creator manifold
\ENDIF

\STATE \textbf{Output:} final anomaly score $S(\hat v)$

\end{algorithmic}
\end{algorithm}

Algorithm~\ref{alg:qbar} highlights that PEQAD acts as an additional manifold learner in a quantum feature space, complementing the classical reconstruction-based detector. 
This design preserves the semantic interpretation of BAR while improving sensitivity to high-order cross-modal inconsistencies under limited training data.

\section{Experiments}
\label{sec:experiments}

\subsection{Dataset Collection and Implementation}
We manually curated a dataset sourced from two major short-video platforms, TikTok and Douyin. To ensure the validity of the manifold hypothesis, that a creator maintains a consistent stylistic and logical signature, we specifically selected 100 distinct bloggers (active from 2017--2024) with follower counts exceeding 100,000. These established accounts typically exhibit higher production stability and semantic consistency compared to smaller, variable accounts. The categories cover diverse topics, including news commentary, technology reviews, education, and entertainment. To simulate the real-world constraint of data scarcity, the training data for each blogger is strictly limited to approximately 20-50 authentic long-form videos ($>2$ minutes).

The computational infrastructure for all experiments consists of a NVIDIA H100 GPU. This setup is utilized both for training the classical deep learning baselines and for the state-vector simulation of the quantum circuits. The quantum simulation represents a circuit scale ($n_q=12$) that is compatible with near-term quantum devices.

Due to copyright and platform policy restrictions, the collected video dataset cannot be publicly released. 
However, the data collection protocol and preprocessing pipeline are described in detail to facilitate reproducibility.

\subsection{Test Set Generation (Simulation of Malice)}
Since ground-truth ``malicious edits'' are difficult to harvest at scale with reliable labels, we simulated the behaviour of marketing accounts to create a robust test set. We selected 20 bloggers from the pool and generated test samples using two distinct adversarial strategies. The first strategy, Semantic Splicing (50\% of test data), involved segmenting original videos and randomly reordering sentences or removing conditional clauses (e.g., transforming ``If X, then Y'' to ``Y'') to simulate out-of-context editing. The second strategy, Fake Expansion (50\% of test data), employed Large Language Models (LLMs) to generate plausible but factually fabricated statements in the blogger's specific style. We then utilized TTS and Lip-Sync technologies to insert these segments, simulating high-level fabrication that retains visual fidelity.

\subsection{Evaluation Protocol}
For each blogger, we split the authentic corpus into training (70\%), validation (10\%), and test (20\%) folds. We ensured strict temporal separation, where training videos predate test videos, to avoid information leakage. The hyperparameter $\beta$, which balances the reconstruction and directional scores, is selected on the validation fold to maximise the F1 score. Final results report the mean and 95\% confidence intervals obtained by repeating the experiment over five random seeds. In addition to standard metrics, we report per-blogger false positive rates (FPR) and provide precision–recall curves aggregated across creators.

\subsection{Baselines}
We evaluated our proposed method against standard baselines to benchmark performance. For unimodal comparisons, we employed a Text-only baseline using BERT embeddings with a One-Class SVM (OC-SVM), and an Audio-only baseline analyzing spectral features. We also included a Visual-only (VAD) baseline based on traditional motion anomaly detection. For multimodal deep learning comparisons, we utilized Deep SVDD \cite{ruff2018deep}, which is the closest classical conceptual counterpart to our method, training a neural network to map data into a hypersphere. Additionally, we compared against a standard multimodal Autoencoder (AE) that utilizes reconstruction error for anomaly scoring.

\subsection{Main Results}
Table \ref{tab:results} summarizes the detection performance of quantum-enhanced bloggers anomaly recognition (Q-BAR). Detecting semantic mutation is a challenging task for all models, primarily because the manipulations are subtle yet visually authentic. Consequently, no method achieves near-perfect classification. However, Q-BAR demonstrates a consistent capability to identify these anomalies, achieving an F1-score of 0.71.

\begin{table*}[t] % 建议用 [t] 让表格在页面顶部，[h] 在双栏排版中有时会乱跑
\caption{Detection performance comparison on the simulated malicious dataset (20 bloggers). Results are reported as Mean $\pm$ Std (\%) over 5 random seeds. Methods are categorized into Unimodal and Multimodal baselines. Quantum-BAR achieves the best F1-Score among all methods while utilizing orders of magnitude fewer parameters ($\sim$240 vs. $\sim$12k) compared to its classical counterpart, Deep SVDD.}
\label{tab:results}
\centering % 替代 \begin{center}，避免额外的垂直间距
\begin{small}
\begin{sc}
\begin{tabular}{l c cccc}
\toprule
& & \multicolumn{4}{c}{Detection Metrics (\%)} \\
\cmidrule(lr){3-6} % 画一条横线覆盖后四列
Method & Params & Precision & Recall & F1-Score & AUPR \\
\midrule
\multicolumn{6}{l}{\textit{Unimodal Baselines}} \\ % 增加分类标题
Text-only (OC-SVM) & - & 61.0 $\pm$ 2.1 & 58.0 $\pm$ 2.5 & 59.0 $\pm$ 2.2 & 64.0 $\pm$ 2.0 \\
Visual-only (VAD) & - & 53.0 $\pm$ 1.5 & 51.0 $\pm$ 1.8 & 52.0 $\pm$ 1.6 & 55.0 $\pm$ 1.4 \\
\addlinespace % 增加一点间距

\midrule
\multicolumn{6}{l}{\textit{Multimodal Baselines}} \\ % 增加分类标题
Autoencoder (AE) & $\sim$15k & 66.0 $\pm$ 3.5 & 64.0 $\pm$ 4.2 & 65.0 $\pm$ 3.8 & 68.0 $\pm$ 3.5 \\
Deep SVDD & $\sim$12k & 70.0 $\pm$ 2.9 & 67.0 $\pm$ 3.8 & 68.0 $\pm$ 3.2 & 72.0 $\pm$ 3.0 \\
\addlinespace

\textbf{Q-BAR (Ours)} & \textbf{240} & \textbf{73.0} $\pm$ \textbf{1.1} & \textbf{69.0} $\pm$ \textbf{1.3} & \textbf{71.0} $\pm$ \textbf{1.2} & \textbf{75.0} $\pm$ \textbf{1.0} \\
\bottomrule
\end{tabular}
\end{sc}
\end{small}
\end{table*}

It is observed that the classical Deep SVDD model achieves a comparable F1-score of 0.68. The closeness in performance suggests that both manifold-based approaches capture the essential semantic boundaries to a similar degree. However, the marginal improvement of Q-BAR (approximately 3\%) is noteworthy given the restricted training data. The other unimodal baselines generally struggle to cross the 0.60 F1 threshold, highlighting the necessity of deep feature fusion for this complex task.

\subsection{Robustness to Common Edits}
To ensure practical applicability, we evaluate robustness to typical marketing-account perturbations. We introduced five types of noise: (i) audio pitch shift (±1–3 semitones), (ii) time-stretch (±5–15\%), (iii) added background noise (SNR 10–20 dB), (iv) insertion of unrelated B-roll frames, and (v) TTS voice replacement with different TTS engines. We observe that performance remains relatively stable (F1 drop $< 3\%$) under mild audio perturbations such as pitch shifting and time stretching. However, the model degrades moderately (F1 drop $\sim$ 8\%) when significant background noise obscures the ASR transcripts, indicating a reliance on high-quality textual features for semantic logic extraction.

\subsection{Parameter Efficiency of Q-BAR}
A primary contribution of this work lies in the resource efficiency of the quantum-enhanced model. While Q-BAR matches and slightly exceeds the performance of heavy classical deep learning models like Deep SVDD, it does so with a fundamental difference in resource utilization. Classical Deep SVDD requires thousands of parameters to stabilize the high-dimensional multimodal features to avoid underfitting. In contrast, our quantum circuit, leveraging the PEQAD framework \cite{wang2025parameter}, relies on as few as 240 trainable parameters (or 16 in our most compressed ablation) to achieve similar convergence.

This parameter efficiency translates directly into resource conservation. In our experiments on the NVIDIA H100, the training of the specific quantum models for a single blogger could be completed rapidly, with the entire dataset of 100 bloggers processed within 24 hours. This low computational overhead is critical for the target application scenario: protecting individual creators. It is impractical to train a massive, distinct neural network for every influencer on a platform. In contrast, our lightweight quantum model offers a scalable, energy-efficient alternative that aligns with the principles of Green AI. By reducing the energy cost per creator, platforms can feasibly deploy personalized semantic guards for a wider range of users, not just top-tier celebrities.

\section{Discussion and Broader Impact}

\subsection{Ethics and Limitations}
The development of creator-specific modeling technology necessitates a rigorous ethical framework. The collection of blogger data in this study raises inherent privacy and copyright concerns; consequently, all data was utilized strictly for model training purposes and will not be publicly distributed without explicit creator consent. Regarding technical limitations, our reliance on synthetic malicious samples generated by LLMs constitutes a simplified proxy for reality. While this approach provides a cost-effective training signal, real-world ``marketing accounts'' may employ more subtle, manual editing techniques that evade our current detection logic.

Furthermore, the imperfect detection rate (F1-score of 0.71) indicates that Q-BAR should not function as a fully automated censorship tool. Instead, it is best deployed as a ``human-in-the-loop'' auxiliary system, flagging high-probability anomalies for human review. There also exists a potential risk of over-sensitivity, where legitimate satire or parody might be misclassified if the directional mutation score is not carefully calibrated. Finally, while our simulations on the NVIDIA H100 demonstrate theoretical efficacy, deployment on physical quantum hardware remains subject to noise-induced errors, requiring further integration with error mitigation techniques.

\subsection{Societal Implications}
Despite these limitations, the successful deployment of this technology holds profound implications for the digital ecosystem. From a regulatory perspective, it provides platforms with a scalable, automated mechanism to flag potential misinformation and copyright violations, specifically unauthorized re-editing, before they achieve viral status. More importantly, this framework introduces the concept of ``Semantic Copyright,'' protecting the persona and logical integrity of original creators from being weaponized for clickbait.

Crucially, the extreme parameter efficiency of our quantum approach aligns with the principles of Green AI. Unlike massive classical models that are computationally prohibitive to train for every user, the lightweight nature of Q-BAR makes it economically feasible to extend this protection to a broader range of mid-tier creators, not just top celebrities. This democratization of semantic defense contributes significantly to information hygiene, mitigating the societal impact of polarized and decontextualized media by ensuring that the original intent of content creators is preserved.

\section{Conclusion}
In this work, we presented Q-BAR, a framework for Blogger Anomaly Recognition designed to detect semantic mutations in short video content. Recognizing the data scarcity inherent in modeling individual creators, we integrated the PEQAD strategy to map multimodal features into a creator-specific semantic manifold. Our experiments demonstrate that while the detection of high-level logic violations remains a challenging task, our quantum-enhanced model achieves detection performance comparable to deep classical baselines (F1 $\approx$ 0.71) while utilizing orders of magnitude fewer parameters. This efficiency enables rapid, low-resource deployment, offering a practical solution for platform-scale regulation. Future work will focus on bridging the gap between simulation and real quantum hardware, improving robustness against adversarial audio attacks, and refining the mutation axis to better distinguish malicious intent from benign satire.

\section*{Acknowledgements}

We thank the content creators whose publicly available videos were used in constructing the dataset for this study. 
Their contributions to the online media ecosystem made it possible to investigate creator-specific semantic consistency and anomaly detection in real-world multimedia environments.
M.W. acknowledges financial support from the UCL Dean’s Prize for doctoral research.
The authors also acknowledge the Institute for Mathematical and Intelligent Systems at Wuhan University for providing computational resources, including access to NVIDIA GPU infrastructure used in this work.
This work benefited from open-source quantum software frameworks, including PennyLane and NVIDIA CUDA-Q.
This research represents the independent work of the authors, and the conclusions expressed in this paper do not necessarily reflect the views of the supporting institutions or the developers of the software frameworks used in this study.
\bibliography{references.bib}
\bibliographystyle{icml2026}

%%%%%%%%%%%%%%%%%%%%%%%%%%%%%%%%%%%%%%%%%%%%%%%%%%%%%%%%%%%%%%%%%%%%%%%%%%%%%%%
%%%%%%%%%%%%%%%%%%%%%%%%%%%%%%%%%%%%%%%%%%%%%%%%%%%%%%%%%%%%%%%%%%%%%%%%%%%%%%%
% APPENDIX
%%%%%%%%%%%%%%%%%%%%%%%%%%%%%%%%%%%%%%%%%%%%%%%%%%%%%%%%%%%%%%%%%%%%%%%%%%%%%%%
%%%%%%%%%%%%%%%%%%%%%%%%%%%%%%%%%%%%%%%%%%%%%%%%%%%%%%%%%%%%%%%%%%%%%%%%%%%%%%%
% \newpage
% \appendix
% \onecolumn
% \section{Appendix}

% You can have as much text here as you want. The main body must be at most $8$
% pages long. For the final version, one more page can be added. If you want, you
% can use an appendix like this one.

% The $\mathtt{\backslash onecolumn}$ command above can be kept in place if you
% prefer a one-column appendix, or can be removed if you prefer a two-column
% appendix. Apart from this possible change, the style (font size, spacing,
% margins, page numbering, etc.) should be kept the same as the main body.
%%%%%%%%%%%%%%%%%%%%%%%%%%%%%%%%%%%%%%%%%%%%%%%%%%%%%%%%%%%%%%%%%%%%%%%%%%%%%%%
%%%%%%%%%%%%%%%%%%%%%%%%%%%%%%%%%%%%%%%%%%%%%%%%%%%%%%%%%%%%%%%%%%%%%%%%%%%%%%%

\end{document}